\documentclass[a4paper,12pt]{article}

\usepackage{amssymb,amsmath}
\usepackage[totalwidth=17cm,totalheight=24cm]{geometry}
\usepackage{graphicx}

\newcommand{\R}{{\mathbb R}}

\newcommand{\id}{{\mathbb I}}

\newcommand{\be}{\begin{eqnarray}}
\newcommand{\ee}{\end{eqnarray}}

\begin{document}
 \pagestyle{plain}
\title{Field redefinitions and Plebanski formalism for GR}
\author{Kirill Krasnov \\ {\it School of Mathematical Sciences, University of Nottingham, NG7 2RD, UK}}

\date{August 2017}
\maketitle

\begin{abstract}\noindent We point out that there exists a family of transformations acting on $BF$-type Lagrangians of gravity, with Lagrangians related by such a transformation corresponding to classically equivalent theories. A transformation of this type corresponds to a particular field redefinition. We discuss both the chiral and non-chiral cases. In the chiral case there is a one-parameter, and in the non-chiral case a two-parameter family of such transformations. In the chiral setup, we use these transformations to give an alternative derivation of the chiral $BF$ plus potential formulation of General Relativity that was proposed recently. In the non-chiral case, we show that there is a new $BF$ plus potential type formulation of GR. We also make some remarks on the non-chiral pure connection formulation.

\end{abstract}

\section{Introduction}

In \cite{Plebanski:1977zz} Plebanski proposed a formalism for General Relativity (GR) that is based on only one of the two chiral halves of the spin connection of a Riemannian (pseudo-Riemannian) four-dimensional manifold. This formalism was rediscovered in \cite{Capovilla:1991qb} as giving the most natural Lagrangian realisation of the new Hamiltonian formulation \cite{Ashtekar:1987gu} of GR. In Plebanski formalism GR takes the form of a constrained BF theory. The paper \cite{Krasnov:2009pu} gives a review of this formalism from a practical viewpoint. In particular, it is explained how it makes considerably simpler the procedure of arriving at Einstein's equations for a given 4D metric. Another useful source is \cite{Freidel:2012np}, which also discussed the non-chiral version.

General Relativity in Plebanski formalism can be modified in a very interesting way. These modifications were first discovered in the context of "pure connection" formulation \cite{Capovilla:1989ac}. This paper pointed out the existence of a one-parameter family of modifications. The works \cite{Capovilla:1992ep} and \cite{Bengtsson:1990qh} further studied this theory. Bengtsson \cite{Bengtsson:1990qg} was the first to realise the existence of an infinite parameter family of modifications. He dubbed these theories "neighbours of GR" and studied them intensively, see \cite{Bengtsson:1991bq}-\cite{Bengtsson:1995tx}. What is very interesting about these 4D modified gravity theories is that they continue to propagate two degrees of freedom, as General Relativity. 

It is important to remark that these theories provide a modification of complexified General Relativity. Reality conditions should then be imposed to select an appropriate real slice. While it is easy to select real slices corresponding to Euclidean or split signatures, no reality condition is known in the modified case that would be appropriate for the Lorentzian signature. This was one of the problems tackled by Bengtsson in the early days of the studies of these theories, and this status quo persisted till now. In this paper, when we refer to modified gravity theories of this type, we imply the Euclidean version of the formalism. 

Bengtsson studied these modified gravity theories mostly from the pure connection or Hamiltonian viewpoints. These theories were rediscovered by the present author \cite{Krasnov:2006du} in a BF-type formalism. The relation to previous work was pointed out in \cite{Bengtsson:2007zx}. 

In retrospect, one of the easiest ways to explain what these modified theories are is to start with the Plebanski BF-type action for GR, and replace the constraint term for the 2-form $B$-field with a potential term, see \cite{Krasnov:2008fm} and also \cite{Krasnov:2009iy}. The interpretation proposed in \cite{Krasnov:2008fm} was that when the potential becomes "infinitely steep" one recovers GR. For arbitrary potentials one gets modified theories. Another way of saying the same thing is that General Relativity arises as the low energy limit of any of the modified theories, and so modifications we are talking about are UV modifications. These theories were referred to as "deformations of GR" by the present author. 


The main motivation for the constructions of this paper comes from the recent work \cite{Herfray:2015rja} that proposed a new BF-type Lagrangian formulation of (unmodified) General Relativity. In this new formulation there are no Plebanski-type simplicity constraints imposed on the 2-form field. The Lagrangian is just a functional of the connection and the 2-form field. 

In retrospect, the most surprising fact about the new formulation \cite{Herfray:2015rja} is that it puts GR into the form of one of theories from the "deformations of GR" class. It is a BF-type theory with a potential for the $B$-field. And at the same time the claim is that it continues to describe the usual General Relativity. How is this possible given the suggested above interpretation that GR corresponds to an "infinitely steep" potential?

This is the puzzle that we are going to address in the present paper. Our main observation is that there is a one-parameter family of field redefinitions that one can perform on the Plebanski action of GR, with the result of this field redefinition being the formulation discovered in \cite{Herfray:2015rja}. What this means is that the interpretation of GR as corresponding to an infinitely steep potential should be revisited. The situation is more interesting, as we will describe below. 

The paper is organised as follows. We start in Section \ref{sec:Pleb} by a review of the chiral Plebanski formalism, and modified gravity theories that can be naturally obtained in this framework. Section \ref{sec:field-redef} is the central one. We introduce the field redefinitions here, and derive the main result of this paper. Section \ref{sec:chiral} describes the first application of the main result. Thus, we re-derive the formulation proposed in \cite{Herfray:2015rja} by what appears to be a simpler argument. We turn to non-chiral $BF$-type formalism for GR in Section \ref{sec:non-chiral}. We show that our results imply that there exists a previously unknown $BF$ plus potential non-chiral description of GR. We also discuss here the non-chiral pure connection formulation, and explain why it is difficult to write down a useful closed-form action in this case. We conclude with a discussion.

\section{Review of Plebanski formalism and "deformations"}
\label{sec:Pleb}

\subsection{Plebanski formulation of General Relativity}

We start with a brief review of the Plebanski formalism. The action is of the so-called BF-type, with a Lagrange multiplier term imposing a constraint on the $B$-field. It reads
\be\label{action-Pleb}
S_{\rm Pleb}[B,A,\Psi] = \int B^i F^i - \frac{1}{2} \left( \Psi^{ij} + \frac{\Lambda}{3} \delta^{ij}\right) B^i B^j.
\ee
Here $B^i$ is a 2-form field taking values in the Lie algebra of ${\rm SO}(3)$, with $i=1,2,3$ being the Lie algebra index. $F^i = dA^i + (1/2) \epsilon^{ijk} A^j A^k$ is the curvature of an ${\rm SO}(3)$ connection $A^i$. The field $\Psi^{ij}$ is referred to as the Lagrange multiplier field. It is assumed to be trace-free ${\rm Tr}(\Psi)\equiv \Psi^{ij}\delta_{ij}=0$. The quantity $\Lambda$ is the cosmological constant. Wedge product of differential forms is assumed everywhere. Constants in front of the action, such as Newton's constant and factors of $\pi$ are not relevant to our discussion and omitted. 

Varying with respect to the Lagrange multiplier field one obtains constraints on the $B$-field
\be\label{simplicity}
B^i B^j \sim \delta^{ij}.
\ee
These are known under the name of "simplicity constraints" in the literature. These constraints can be shown to imply that the 2-form field "comes from a tetrad", i.e. can be written as the self-dual projection of the wedge product of two tetrads. Detailed explanation and proof of this statement can be found in the previously cited literature on Plebanski formalism.

Varying with respect to the connection one gets the equation that states
\be
d_A B^i =0,
\ee
where $d_A$ stands for the exterior covariant derivative. This equation, together with (\ref{simplicity}), then implies that $A^i$ is the self-dual part of the spin connection compatible with the tetrad that was introduced in the previous paragraph. 

Varying with respect to the 2-form field one then gets the Plebanski formalism version of Einstein equations
\be\label{F-B}
F^i = \left( \Psi^{ij} + \frac{\Lambda}{3} \delta^{ij}\right) B^j.
\ee
This set of equations say that the curvature of the self-dual part of the spin connection is self-dual as the 2-form. This is known to be equivalent to the Einstein condition. This equation also interprets the on-shell value of the Lagrange multiplier field $\Psi^{ij}$ as the self-dual part of the Weyl curvature tensor. For more details on this formalism we refer the reader to e.g. \cite{Krasnov:2009pu}.

\subsection{"Deformations of GR"}

As we have described in the Introduction, one way to think about "deformed GR" theories is to replace the Lagrange multiplier term in (\ref{action-Pleb}) with a general potential for the 2-form field. However, it was later realised that there is an even simpler way to introduce the modified theories. It is based on an obvious, but what turns out to be powerful rewriting of the Plebanski action
\be\label{action-mu}
S[B,A,M,\mu] = \int B^i F^i - \frac{1}{2} M^{ij} B^i B^j + \mu ({\rm Tr}(M) - \Lambda).
\ee
Thus, we have added a new Lagrange multiplier field $\mu$ that explicitly imposes the constraint that the trace of the matrix $M^{ij}$ that appears in front of the wedge product of 2-form fields is constant. It is clear that (\ref{action-mu}) is equivalent to (\ref{action-Pleb}). 

The deformations of GR can now be introduced as theories obtained by deforming the constraint in (\ref{action-mu}). Indeed, let us write instead of ${\rm Tr}(M)$ an arbitrary ${\rm SO}(3)$-invariant function of the matrix $M^{ij}$
\be\label{action-def}
S[B,A,M,\mu] = \int B^i F^i - \frac{1}{2} M^{ij} B^i B^j + \mu (f(M) - \Lambda).
\ee
One can obtain some very interesting theories by changing $f(M)$. The Plebanski case is
\be\label{f-GR}
f_{\rm GR}(M) = {\rm Tr}(M).
\ee
It turns out that there is also a choice that gives the theory of gravitational instantons
\be
f_{\rm inst}(M) = {\rm Tr}(M^{-1}).
\ee
This fact is explained in \cite{Herfray:2015fpa}. This reference also puts to full use the parametrisation (\ref{action-def}) of the deformed theories. It turns out that it is very convenient for solving the field equations. 

The parametrisation (\ref{action-def}) is also a convenient step towards the "pure connection" formulation of this class of gravity theories, and of General Relativity in particular. Indeed, both the 2-form field and the auxiliary matrix field $M$ can be integrated out, with what results being the pure connection action. When applied to the case (\ref{f-GR}) this procedure yields the pure connection action \cite{Krasnov:2011pp}, see \cite{Celada:2015jda} for details of this derivation. 

We finally remark that one can also integrate out from (\ref{action-def}) just the auxiliary matrix field $M$, with what results being the BF plus potential type formulation of the deformed theories. This is not possible for the case of General Relativity though because $\partial f_{\rm GR}/\partial M=const$. Thus, in the case of GR one cannot use the Euler-Lagrange equation for $M$ to solve for $M$. This was one of the reasons why GR was interpreted in \cite{Krasnov:2008fm} as corresponding to an infinitely steep potential. But the present paper will revise this interpretation. 

\section{Field redefinitions}
\label{sec:field-redef}

We now come to the main point of this paper and consider a certain class of field redefinitions that can be performed on actions of the type (\ref{action-def}). Thus, consider the transformation
\be\label{transform}
B^i = G^{ij} \tilde{B}^j + H^{ij} F^j,
\ee
where $G^{ij}, H^{ij}$ are arbitrary at this stage $3\times 3$ matrices, and $\tilde{B}$ is the new 2-form field. This transformation will map the first two terms in the Lagrangian (\ref{action-def}) to
\be
L \to \tilde{B}^{tr} G^{tr} F + F^{tr} H^{tr} F - \frac{1}{2}(\tilde{B}^{tr} G^{tr} + F^{tr} H^{tr}) M (G \tilde{B} + H F),
\ee
where we used the matrix notations with e.g. $M^{ij}B^i B^j \equiv B^{tr} M B$. Collecting the similar terms in the above expression we rewrite it as
\be\label{L-transf}
L\to F^{tr} \left( H^{tr} - \frac{1}{2} H^{tr} M H\right) F + \tilde{B}^{tr} \left( G^{tr} - G^{tr} MH \right) F - \frac{1}{2} \tilde{B}^{tr} (G^{tr} M G) \tilde{B}.
\ee
We now demand that after the transformation (\ref{transform}) the Lagrangian is still of BF-type, i.e. the matrix appearing in front of $\tilde{B}^i F^j$ is a multiple of the identity matrix. If we don't want to change the coefficient in front of the action, we should demand this multiple to be unity
\be\label{eq-G}
G^{tr} - G^{tr} MH  = \id.
\ee
We will also demand that the newly generated term quadratic in the curvature is a multiple of the Pontryagin number for the ${\rm SO}(3)$ bundle in question. Thus, we demand that also the matrix in front of $F^i F^j$ is a constant multiple of the identity
\be\label{eq-H}
H^{tr} - \frac{1}{2} H^{tr} M H = t \id,
\ee
where $t$ is an arbitrary parameter, real if we specialise the formalism to the cases of Euclidean or split signatures. 

We are now going to solve the equations (\ref{eq-H}) and (\ref{eq-G})  for $H,G$ in terms of $M$. First, the equation (\ref{eq-H}) tells us that $H$ is a symmetric matrix, so will will drop the transpose symbol on $H$ from now on. Assuming that $G,H$ are invertible, we can rewrite the two equations (\ref{eq-G}), (\ref{eq-H}) as
\be\label{eqs-GH}
\id - MH = (G^{tr})^{-1}, \qquad \id - \frac{1}{2} MH = t H^{-1}.
\ee
We can then subtract twice the second equation from the first to get a relation between $G,H$
\be\label{H}
H = 2 t (\id + (G^{tr})^{-1})^{-1},
\ee
where we again assumed that $\id + (G^{tr})^{-1}$ is invertible. We then substitute this to e.g. the first equation in (\ref{eqs-GH}) to obtain a simple equation involving just $G$
\be
(G^{tr})^{-2} = \id - 2t M.
\ee
This tell us that $G$ is also a symmetric matrix, and gives this matrix as one of the two branches of the square root
\be\label{G}
G= \left( \id - 2t M\right)^{-1/2}.
\ee

We can now concentrate on the last $BB$ term in (\ref{L-transf}). It is clear that the matrix in front of  $B^i B^j$ transforms to
\be
\tilde{M} = M\left( \id - 2t M\right)^{-1}.
\ee
We note for future use that the inverse of this is $M = \tilde{M}\left( \id + 2t \tilde{M}\right)^{-1}$. 

All in all we learn that the field redefinition (\ref{transform}) with symmetric matrices $G,H$ that depend on $M$ according to (\ref{G}), (\ref{H}) transform the Lagrangian in (\ref{action-def}) into a Lagrangian of the same type
\be
B^i F^i - \frac{1}{2} M^{ij} B^i B^j + \mu (f(M) - \Lambda) = \\ \nonumber
\tilde{B}^i F^i - \frac{1}{2} \tilde{M}^{ij} \tilde{B}^i \tilde{B}^j + \mu (f\left(\tilde{M}( \id + 2t \tilde{M})^{-1}\right) - \Lambda) + t F^i F^i.
\ee
The only change in the new Lagrangian is that the function $f(M)$ became modified, and that a constant multiple of the topological term ${\rm Tr}(F\wedge F)$ has been added. 

Thus, we learn that there is a one-parameter group of transformations acting on the space of theories of the type (\ref{action-def}), with all functions $f(M)$ belonging to the family
\be\label{f-t}
f_t(M) \equiv f(M_t), \qquad M_t = M (\id + 2t M)^{-1}
\ee
corresponding to (classically) physically equivalent theories. At the quantum level adding to the Lagrangian a topological term is not innocuous, as the example of the $\theta$-term in QCD teaches us. So, we can only be sure about the classical equivalence of theories related by (\ref{f-t}). Note that we can alternatively write $M_t^{-1} = M^{-1} + 2t\id$, from which he fact that the transformation $M\to M_t$ forms a one-parameter group $(M_{t_1})_{t_2} = M_{t_1+t_2}$ is obvious. 

The result (\ref{f-t}) can be given a simpler derivation at the level of pure connection formalism. We shall describe this in the last section. 

The equivalence of theories with defining functions related as in (\ref{f-t}) is the main result of this section. We now turn to applications.

\section{GR as $BF$ theory plus potential}
\label{sec:chiral}

We now use the result (\ref{f-t}) to re-derive the new formulation of GR discovered in \cite{Herfray:2015rja}. Logically the simplest way of doing this is to integrate out the auxiliary matrix $M$ from the Lagrangian (\ref{action-def}) with the defining function (\ref{f-t}). 

To this end it is convenient to write 
\be\label{XB}
\frac{1}{2} B^i\wedge B^j = \tilde{X}_B^{ij} d^4x,
\ee
where $\tilde{X}_B^{ij}$ is a density weight one matrix and $d^4x$ is the coordinate volume form in some arbitrary coordinate system. We also write the Lagrange multiplier $\mu=\tilde{\mu} d^4x$. The equation for $M$ is then
\be\label{XB-M}
\tilde{X}_B^{ij} = \tilde{\mu} \frac{\partial f}{\partial M}.
\ee
For the defining function $f_t$ we have
\be
 \frac{\partial f_t}{\partial M} = (\id + 2t M)^{-2},
 \ee
and so we can solve (\ref{XB-M}) for $M$ 
\be
2t M = \sqrt{\tilde{\mu}} (\tilde{X}_B)^{-1/2} - \id,
\ee
where we assumed that $\tilde{X}_B$ is invertible and one of the two branches of the square root is taken. This gives
\be
M(\id + 2t M)^{-1} = \frac{1}{2t}\left(\id - \sqrt{ \frac{\tilde{X}_B}{\tilde{\mu}}} \right).
\ee
We should now find $\tilde{\mu}$ from the constraint that the trace of the matrix above is $\Lambda$. This gives
\be
\sqrt{\tilde{\mu}} = \frac{ {\rm Tr}\sqrt{ \tilde{X}_B}}{3-2t \Lambda},
\ee
and so
\be
M = \frac{1}{2t} \left( \frac{ {\rm Tr}\sqrt{ \tilde{X}_B}}{3-2t \Lambda} (\tilde{X}_B)^{-1/2} - \id\right).
\ee
The sought potential is then
\be
V(\tilde{X}_B) \equiv {\rm Tr}(M \tilde{X}_B) = \frac{1}{2t} \left( \frac{ \left({\rm Tr}\sqrt{ \tilde{X}_B}\right)^2}{3-2t \Lambda}  - {\rm Tr}(\tilde{X}_B) \right).
\ee
All in all, returning to using the 2-form field $B^i$, we can rewrite the action (\ref{action-def}) with the defining function (\ref{f-t}) and with the matrix $M$ integrated out as
\be\label{action-GR-BF}
S[B,A]= \int B^i F^i - \frac{1}{4t(3-2t\Lambda)} \left({\rm Tr}\sqrt{ B^i B^j}\right)^2  +\frac{1}{4t}  B^i B^i,
\ee
where the meaning of the first expression in the potential should be interpreted by passing to $\tilde{X}_B$ via (\ref{XB}). The action (\ref{action-GR-BF}) is just of the type that was proposed in \cite{Herfray:2015rja} as giving a description of GR. We have thus confirmed that (\ref{action-GR-BF}) is obtained from the Plebanski action for GR by a field redefinition and by the procedure of integrating out the auxiliary field. It thus gives a classically equivalent description of GR. In \cite{Herfray:2015rja} this was demonstrated by analysing the field equations following from (\ref{action-GR-BF}), but the above derivation using the field redefinitions gives a more transparent argument. 

It is interesting to note that while one is unable to integrate out the matrix $M$ from the GR action (\ref{action-mu}), it becomes possible to do so starting from the action after the field redefinition (\ref{transform}) is applied. 

This is a good place to discuss the effect of the field redefinition (\ref{transform}) on the metric. On-shell the curvature 2-forms become linear combinations of the 2-forms $B^i$. This is true in the case of GR, see (\ref{F-B}), as well as for the modified theories. Because the conformal class of the metric is fixed by demanding that the 2-forms $B^i$ span the space of self-dual 2-forms, the conformal class is unchanged by the field redefinition (\ref{transform}). However, this transformation does have the effect on the volume form that fixes a representative in the conformal class. In particular, the volume form which corresponds to an Einstein metric is constructed differently in the Plebanski case and the formulation (\ref{action-GR-BF}). This is explained in more details in \cite{Herfray:2015rja}.

The action (\ref{action-def}) with (\ref{f-t}) can also be used as the starting point for deriving the pure connection formulation. This is obtained by first integrating out the 2-form field, and then integrating out the auxiliary matrix $M$. We leave it as an exercise to the reared to check that this reproduces the pure connection GR Lagrangian of \cite{Krasnov:2011pp} plus a multiple of the topological term $F^i F^i$, as it should. We also discuss the derivation of the pure connection action for a general member of the modified family of theories in the last section. 

\section{New non-chiral $BF$-type Lagrangian for GR}
\label{sec:non-chiral}

\subsection{Non-chiral Plebanski Lagrangian}

We now come to an application of the above ideas that leads to a previously unknown non-chiral formulation of General Relativity. To this end, we must start by reviewing the non-chiral version of the Plebanski Lagrangian, first analysed in details in \cite{DePietri:1998hnx}. The gauge group that plays role in this formulation is the full Lorentz group. Let $I,J,\ldots = 1,2,3,4$ denote the indices for $\R^4$ or $\R^{1,3}$, so that objects of the type $X^{IJ}=X^{[IJ]}$ are in the Lie algebra of the Lorentz group (of appropriate signature). The non-chiral Plebanski action is
\be\label{action-non-chiral}
S[B,A,\Psi] = \int B_{IJ} F^{IJ} - \frac{1}{2} \left( \Psi^{IJKL} + \frac{\Lambda}{24} \epsilon^{IJKL}\right) B_{IJ} B_{KL}.
\ee
Here $B^{IJ}$ is a Lie algebra valued 2-form field, $F^{IJ}$ is the curvature of the spin connection $A^{IJ}$, $\Psi^{IJKL}$ is the Lagrange multiplier field that is assumed tracefree $\Psi^{IJKL}\epsilon_{IJKL}=0$, and $\Lambda$ is a multiple of the cosmological constant. 

Variation with respect to the Lagrange multiplier field gives a set of equations on the 2-form field $B^{IJ}B^{KL} \sim \epsilon^{IJKL}$. As is shown in \cite{DePietri:1998hnx}, these imply that the 2-form field is either the wedge product of two tetrads, or the dual of such wedge product. When one of these two sectors of solutions is substituted into the action (\ref{action-non-chiral}) one recovers the Einstein-Cartan action in terms of the frame and the spin connection, which explains why (\ref{action-non-chiral}) describes General Relativity.

As in the chiral formalism it is useful to rewrite the action (\ref{action-non-chiral}) by explicitly introducing a Lagrange multiplier imposing the trace constraint on the matrix that appears in front of the $B^{IJ} B^{KL}$
\be\label{action-NC-mu}
S[B,A,M,\mu] = \int B_{IJ} F^{IJ} - \frac{1}{2} M^{IJKL} B_{IJ} B_{KL} + \mu ( {\rm Tr}(M) - \sigma \Lambda),
\ee
where the trace that recovers (\ref{action-non-chiral}) is ${\rm Tr}(M)=M^{IJKL} \epsilon_{IJKL}$, and $\sigma=\pm 1$ is the sign that depends on the signature. Minus sign corresponds to the Lorentzian signature. 

As was first pointed out in \cite{Capovilla:2001zi}, it is natural to allow a more general version of the above trace condition. Indeed, the Lorentz group is not simple, and there are two independent invariant bilinear forms on its Lie algebra. It is natural to consider an arbitrary linear combination of the two. The solution of the resulting simplicity constraint is then a linear combination of the wedge product of two tetrads, and the dual of this wedge product. The degeneracy present in the formulation (\ref{action-non-chiral}) is thus removed, and the resulting tetrad action is the Holst version \cite{Holst:1995pc} of the Einstein-Cartan theory. 

Motivated by these considerations, we will consider this more general version of the non-chiral Plebanski theory with the trace condition in (\ref{action-NC-mu}) being
\be
{\rm Tr}_{S}(M) = {\rm Tr}(S M) \equiv M^{IJKL} S_{IJKL}, \quad S_{IJKL}:= \frac{s_1}{2} \delta_{I[K} \delta_{L]J} + \frac{s_2}{2} \epsilon_{IJKL}.
\ee
With this version of the trace one of the 3 parameters $s_1,s_2,\Lambda$ in the constraint in (\ref{action-NC-mu}) is superfluous and can be set to unity by simultaneously rescaling all 3 of them. We set $\Lambda=1$ from now on. 

\subsection{Field redefinitions}

We can now apply the same field redefinitions idea to (\ref{action-NC-mu}). Consider
\be
B^{IJ} = G^{IJKL} \tilde{B}_{KL} + H^{IJKL} F_{KL}.
\ee
We then run through all the steps of the derivation in Section \ref{sec:field-redef}, just replacing the ${\rm SO}(3)$ index $i$ with a pair of indices $IJ$ everywhere. We again demand that the $BF$ term is unmodified, so the equation (\ref{eq-G}) is unchanged. The new feature that arises in the non-chiral formalism is that there are two topological quadratic in the curvature terms that can be written. Indeed, both
\be
\int F^{IJ} F_{IJ}, \qquad \int \epsilon^{IJKL} F_{IJ} F_{KL}
\ee
are topological terms that only depend on the bundle chosen. For a compact Riemannian signature manifold, and when $A^{IJ}$ is the metric compatible spin connection, these are linear combinations of the Euler characteristic and torsion of the manifold. For this reason we now allow a more general version of the equation (\ref{eq-H})
\be\label{eq-H-NC}
H^{tr} - \frac{1}{2} H^{tr} M H = T,
\ee
where we introduced a symmetric matrix
\be
T:= t_1 \, \id + t_2 *.
\ee
Here $t_{1,2}$ are real parameters, and $\id_{IJKL} := (1/2) \delta_{I[K}\delta_{L]J}, *_{IJKL}:=(1/2) \epsilon_{IJKL}$. We now solve (\ref{eq-G}), (\ref{eq-H-NC}) in the way similar to what was done before. The new feature that arises is that the operator $*$ does not in general commute with the matrix $M$, so one has to be careful about the order in which all matrices are multiplied. In particular, the matrix $G$ is no longer symmetric. The solutions of these equations are given by
\be
H= 2T (\id + (G^{tr})^{-1})^{-1}, \qquad (G^{tr})^{-2} = \id - 2MT.
\ee
We now have to be careful again because the matrix $(G^{tr})^{-2}$ on the left-hand-side of the second equation is not necessarily symmetric, and so it is not clear how to take the square root to get $G$ itself. However, we can rewrite the right-hand-side of the second equation as
\be
\id - 2MT = T^{-1/2} ( \id - 2T^{1/2} M T^{1/2} ) T^{1/2},
\ee
where we made some choice of a symmetric square root $T^{1/2}$ of the symmetric matrix $T$, and also assumed that $T$ is invertible. We now note that the matrix that appears in the above expression in brackets is symmetric, and so the notion of its square root makes sense. Thus, we can take the square root of $\id - 2MT$ as
\be
(\id - 2MT)^{1/2} = T^{-1/2} ( \id - 2T^{1/2} M T^{1/2} )^{1/2} T^{1/2}.
\ee
This gives the final solution for $G^{tr}, H$
\be
G^{tr} = T^{-1/2} ( \id - 2T^{1/2} M T^{1/2} )^{-1/2} T^{1/2}, \qquad
H= T^{1/2} ( \id +  T^{1/2} M T^{1/2})^{-1} T^{1/2}.
\ee
We note that the expression for $H$ is symmetric, as it should be. Finally, the new matrix $\tilde{M}\equiv G^{tr} M G$ is given by 
\be
\tilde{M}  = T^{-1/2} ( \id - 2T^{1/2} M T^{1/2} )^{-1/2} (T^{1/2} M T^{1/2}) (\id - 2T^{1/2} M T^{1/2} )^{-1/2} T^{-1/2},
\ee
which is manifestly symmetric as it should be. But now the 3 terms in the middle only contain the matrix $T^{1/2} M T^{1/2}$ and the identity matrix, and so they commute. Therefore we can also write
\be
\tilde{M}  = T^{-1/2}  (T^{1/2} M T^{1/2}) (\id - 2T^{1/2} M T^{1/2} )^{-1} T^{-1/2},
\ee
or more compactly
\be
\id + 2 T^{1/2} \tilde{M} T^{1/2} = (\id - 2 T^{1/2} M T^{1/2})^{-1},
\ee
from which the matrix $M$ in terms of $\tilde{M}$ can be explicitly expressed as
\be
M  = \tilde{M} (\id+ 2 T \tilde{M} )^{-1},
\ee
which finally eliminates the square root of $T$ from the expressions, and is just a generalisation of (\ref{f-t}) to matrix-valued $t$. It can be checked that the right-hand-side is symmetric, as it should be, in spite of $T$ and $\tilde{M}$ not commuting. 

All in all, we learn that there is a two-parameter $t_{1,2}$ family of Lagrangians all giving a classically equivalent description of General Relativity. They are all of non-chiral BF-type and can be written as
\be\label{action-GR-new}
S[B,A,M] = \int B^{tr} F - \frac{1}{2} B^{tr} M B + 
\mu \left( {\rm Tr}\Big[S M (\id + 2T M)^{-1} \Big]-1 \right),
\ee
with $T=0$ corresponding to the original Lagrangian (\ref{action-NC-mu}). Note that we have omitted the tildes from all the quantities in the above Lagrangian. 

\subsection{Towards integrating out $M$}

The natural question now is if there exists a $BF$ plus potential for $B$ type non-chiral formulation of GR. This should be obtained by integrating out the auxiliary matrix $M$ from (\ref{action-GR-new}). 

To carry out such integration, let us introduce as in the chiral case
\be
\frac{1}{2} B^{IJ} B^{KL} = \tilde{X}_B^{IJKL} d^4x,
\ee
and also $\mu = \tilde{\mu} d^4x$. The equation obtained by varying the action with respect to $M$ reads
\be\label{eqn-M}
\tilde{X}_B = \tilde{\mu} (1+2TM)^{-1} S ((1+2TM)^{-1})^{tr}.
\ee
We wrote the equation in a way which makes it manifest that also its right-hand-side is symmetric. 

In the chiral case all the matrices that appeared on the right-hand-side commuted, and it was easy to get the solution for $M$ in terms of some branch of the square root of the symmetric matrix $\tilde{X}_B$. Now the situation is different. One possible way to solve the above equation is to assume that $\tilde{X}_B$ is a positive definite matrix, and use the so-called Cholesky decomposition $\tilde{X}_B= L_B L_B^{tr}$, where $L_B$ is a lower diagonal matrix. Then we can also similarly decompose $S=L_S L_S^{tr}$. In this case the equation (\ref{eqn-M}) implies $L_B = \sqrt{\tilde{\mu}} (1+2TM)^{-1} L_S O$, where $O$ is an orthogonal matrix that needs to be adjusted so that the matrix $M$ obtained from this relation is symmetric. This gives an implicit solution, but it is not clear how much better this implicit solution is than the equation (\ref{eqn-M}) itself, as this equation in principle fixes $M$ in terms of $\tilde{X}_B$. One of course also has to be aware of the fact that there are in general different branches of solutions of this set of quadratic equations on the coefficients of $M$, but this is similar to the chiral case. 

So, there appears to be no useful closed-form expression for the $BF$ plus potential non-chiral GR action, even though it is clear that this action does exist. The situation here is similar to the situation with the pure connection non-chiral action, which exists in principle, but is difficult to write in some explicit useful form. We will review the situation with the pure connection action in the next subsection. 
It appears that the best one can do to characterise the $BF$ plus potential action is to perform its expansion around some appropriate background. For such a characterisation of the non-chiral pure connection action see see \cite{Zinoviev:2005qp}, \cite{Basile:2015jjd}.  And for the purpose of getting a perturbative solution for $M$ around a fixed background the equation (\ref{eqn-M}) is completely adequate.

So, to conclude, we learn that there exists a $BF$ plus potential for the $B$ field non-chiral action for GR, but this action is somewhat difficult to write explicitly, similarly to the situation with the non-chiral pure connection action for GR. This is all in contrast with the situation in the chiral case where both actions can be written explicitly in terms of matrix square roots. 

\subsection{Non-chiral pure connection action revisited}

The purpose of this subsection is to make some comments on the non-chiral pure connection action. This action can be perturbatively solved for starting from the usual frame formalism, see \cite{Zinoviev:2005qp}, \cite{Basile:2015jjd}. The constrained $BF$ formulation of GR (\ref{action-NC-mu}) provides an alternative procedure. However, this alternative also does not lead to any explicit pure connection action, even though may give a simpler set of equations to solve. 

To obtain a pure connection action we start with (\ref{action-NC-mu}), first integrate out the $B$ field, and then integrate out the auxiliary matrix field $M$. We note that even the gauge-theoretic action with $M$ explicitly present may be useful for explicit calculations, as the recent example \cite{Herfray:2015fpa} in the chiral case demonstrates. For the purpose of obtaining the pure connection action one could also start with the new action (\ref{action-GR-new}), but it is easy to see that the result would be the same pure connection action, so one does not get anything new this way.

The action with the $B$ field integrated out reads
\be
S[A,M,\mu] = \int \frac{1}{2} F^{tr} M^{-1} F + \mu ( {\rm Tr}(MS) - 1).
\ee
We now introduce 
\be
\frac{1}{2} F^{IJ} F^{KL} = \tilde{X}_F^{IJKL} d^4x,
\ee
and similarly $\mu = \tilde{\mu}d^4x$. The equation that one needs to solve for $M$ then reads
\be\label{eqn-M-conn}
M^{-1} \tilde{X}_F M^{-1} = \tilde{\mu} S.
\ee
But now, unlike in the chiral case, the matrices $S$ and $M$ do not commute, so we cannot solve this equation in terms of a square root of $\tilde{X}_F$. This is similar to what we have just discussed in the $BF$ formalism. In principle the equation (\ref{eqn-M-conn}) fixes $M$ in terms of $\tilde{X}_F$ and thus gives rise to the pure connection action. But there appears to be no useful explicit way of writing it.

Nevertheless, the equation (\ref{eqn-M-conn}) is a good starting point for getting a perturbative solution for $M$ around some fixed, e.g. constant curvature background. The added benefit of using the $BF$-type formalism for getting the pure connection action, and not the frame Einstein-Cartan formalism, is that in the $BF$ formalism the equation to solve for $M$ is quadratic in $M$, while in the Einstein-Cartan formalism one needs to solve a cubic equation for the frame. Thus, (\ref{eqn-M-conn}) lowers the order of the algebraic equation that needs to be solved to get to the pure connection action, and may be a preferred starting point to get this action. 

We end this subsection by pointing out that the pure connection Lagrangian is given by the Lagrange multiplier field $\mu$, after this is determined from the constraint ${\rm Tr}(MS)=1$. Indeed, we can rewrite the equation (\ref{eqn-M-conn}) as
\be
 \tilde{X}_F M^{-1} = \tilde{\mu} MS,
 \ee
 and then take the trace of both sides. But by the constraint the trace on the right-hand-side gives $\tilde{\mu}$. On the other hand, the trace of the left-hand-side is just our pure connection action we want to compute. So, the pure connection Lagrangian is just the Lagrange multiplier $\mu$, as determined by solving for $M$ from (\ref{eqn-M-conn}) and then substituting into the constraint ${\rm Tr}(MS)=1$. As we already mentioned, this may give a useful procedure for determining the non-chiral pure connection action perturbatively, but we leave this for future work.
 
 \section{Discussion}
 
 This paper showed that there exists a family of transformations acting on the $BF$ plus potential Lagrangians of 4D gravity theories, with Lagrangians related by these transformations corresponding to classically equivalent theories. Such a transformation amounts to a field redefinition of the two-form $B$-field, and adds to the Lagrangian a total derivative term that does not affect the classical theory. There is a one-parameter family of such transformations in the chiral case, and two-parameter family in the non-chiral setup. 

In the chiral case we have used this transformation to give an alternative derivation of the new first-order Lagrangian for GR proposed in \cite{Herfray:2015rja}. We believe that the given here derivation by a field redefinition makes the origin of the new description \cite{Herfray:2015rja} particularly transparent. 

The fact that unmodified General Relativity can be described by the Lagrangian of $BF$ plus potential type, see (\ref{action-GR-BF}), shows that the advocated in \cite{Krasnov:2008fm} interpretation that GR corresponds to an "infinitely steep" potential needs to be revisited. We see that a much better way of characterising the situation is to say that GR corresponds to a specific potential, rather than an "infinitely steep" one. This is true both in the $BF$ plus potential framework, as well as in the pure connection formalism. 

Our results have an implication not just for $BF$ plus potential, but also to the parametrisation in which the family of chiral modified theories was re-discovered in \cite{Krasnov:2006du}. The parametrisation used in that reference was to express the matrix $M$ in terms of its tracefree part $\Psi$ and the trace part $\Lambda$, with $\Lambda=\Lambda(\Psi)$ being a gauge-invariant function of the tracefree part. Thus, the chiral modified theories were described by the Lagrangian of the following type
\be
S[B,A,\Psi] = \int B^i F^i - \frac{1}{2}\left( \Psi^{ij} + \frac{\Lambda(\Psi)}{3} \delta^{ij}\right) B^i B^j.
\ee
The relation to the description (\ref{action-mu}) is to parametrise $M$ in (\ref{action-mu}) as $M^{ij}=\Psi^{ij} + (1/3)\Lambda(\Psi) \delta^{ij}$ and then use the constraint to solve for $\Lambda(\Psi)$. 

It was always assumed that the only function $\Lambda(\Psi)$ that gives GR is a constant function. However, our results show that this is not the case. Indeed, it is clear that the function $\Lambda(\Psi)$ implicitly given by the equation
\be
{\rm Tr}\Big[ \left(\Psi + \frac{1}{3}\Lambda(\Psi)\id\right)\left( (1+ \frac{2t}{3} \Lambda(\Psi)) \id + 2t \Psi \right)^{-1}\Big] = \Lambda
\ee
also describes GR. There is now a one-parameter family of such functions, parametrised by $t$, with only $t=0$ corresponding to $\Lambda(\Psi)=\Lambda$. It is also interesting that the small $t$ expansion of this $\Lambda(\Psi)$ starts with the term ${\rm Tr}(\Psi^2)$, which was always taken as the simplest possible modification. So, we see that there is a way to correct what looks like a modified gravity theory by higher powers of the invariants of $\Psi$ in such a way that there is no modification. We find this amusing. 

Similar remarks apply to the non-chiral case. Thus, the paper \cite{Alexandrov:2008fs} considered a generic modified non-chiral theory, where modification is copied from what it is in the chiral case and consists in allowing the cosmological constant to become a function of the Lagrange multiplier $\Psi$. This work then showed there is in general more propagating degrees of freedom than in the GR case, the maximal number being 8 DOF. Our result (\ref{action-NC-mu}) then shows that there is a choice of the function $\Lambda(\Psi)$ in the non-chiral case that keeps the number of propagating DOF unchanged, and just corresponds to a field redefinition. 

Given that there is a degeneracy in the description of gravity theories in the $BF$ formalism, the question is what is the minimal amount of information that is needed to characterise the classical theory completely. It appears that the pure connection formalism is most efficient in this sense, as the degeneracy of the $BF$ formalism that we have been discussing disappears in the pure connection case. To see this, it is instructive to see how the passage from the $BF$ to the pure connection formalism goes. Let us sketch this in the chiral case, for a general member of the modified family of theories. The action that is an intermediate step between the $BF$ and pure connection formalisms is
\be\label{action-M}
S[A,M,\mu] = \int \frac{1}{2} F^{tr} M^{-1} F + \mu \left( f(M(\id+2tM)^{-1}) - \Lambda\right).
\ee
To integrate out $M$ we need to solve the equation
\be
\tilde{X}_F M^{-2} = \tilde{\mu} f' (1+2tM)^{-2},
\ee
where notations are the same as in the main text, and we use the fact that all matrices appearing here commute. In this equation $f'$ is the derivative of the function $f$ with respect to its argument. This derivative is matrix-valued. Let us denote $M_t=M(\id+2tM)^{-1}$. Then this equation can be rewritten as
\be
\tilde{X}_F (M_t)^{-2}= \tilde{\mu} \frac{\partial f}{\partial M_t}.
\ee
But this is the same equation that would be used for determining the auxiliary matrix starting from the action on which no $t$-transformation has acted. Thus, the only place where the parameter $t$ enters the pure connection Lagrangian is via ${\rm Tr}(M^{-1} \tilde{X}_F)$. But we have $M^{-1} = (M_t)^{-1} - 2t$, which shows that the pure connection Lagrangian obtained from the version of the theory (\ref{action-mu}) with (\ref{f-t}) is the $t=0$ pure connection Lagrangian plus a multiple of the Pontryagin number. This is of course all as expected, because a $B$-field redefinition should play no role once this field is integrated out in the pure connection formalism. We see this explicitly. 

We also note that just described role played by the $t$-transformation in the pure connection formalism can used as an alternative derivation of the fact that all functions (\ref{f-t}) correspond to classically equivalent theories. Indeed, it is clear that they all lead to the same pure connection action, modulo a surface term. However, we believe that the given in the main text derivation via a field redefinition is more transparent. 

This discussion shows that at the level of the pure connection formulation the only degeneracy that exists is the obvious degeneracy of adding the topological term to the action. This once again illustrates the power of the pure connection formulation, because the $t$-transformations act in the space of pure connection Lagrangians in so much simper way. But for practical computations in connection formalism it seems that the intermediate Lagrangian (\ref{action-M}) is more useful than the pure connection one, see \cite{Herfray:2015fpa}. Because of this, one has to be aware of the $t$-degeneracy of the last constraint term in (\ref{action-M}), with different values of $t$ giving classically equivalent theories.  

We also remark that, interestingly, the $t$-transformation action on the space of matrices $M$ via $(M_t)^{-1}=M^{-1} +2t$ is reminiscent of the one-loop YM renormalisation group flow $(g^2_{E'})^{-1} = (g^2_E)^{-1} + c \ln(E'/E)$, where $c$ is a constant that depends on the gauge group, and $E',E$ are two different energy scales. We find this amusing. 

Our final remark is that the non-chiral $BF$-type formalism for GR is the starting point for construction of spin foam models of General Relativity \cite{Perez:2012wv}. The formulation that is usually used for the purpose of constructing spin foam models is the original action (\ref{action-non-chiral}) with constraints imposed on the $B$-field. What this paper showed is that there exists an action of a different type, with no constraints on the 2-form field and no Lagrange multiplier field present. Instead, there is just a potential term for the $B$-field. This action can be obtained by integrating out the matrix $M$ from (\ref{action-GR-new}). This action, while somewhat difficult to write explicitly, does exist, and can, if desired, be computed perturbatively around an appropriate background. This action still describes General Relativity. It is then an interesting challenge for the spin foam framework to explain what kind of spin foam model would correspond to this new action. A related question is how to see the field redefinitions that are the subject of this paper in the discretised context of spin foams. 

\bigskip\noindent{\bf Acknowledgment.} The author is grateful to Yuri Shtanov for reading the manuscript and comments. The author was partially supported by Nottingham Theory Consolidated Grant - ST/L000393/1.

\end{document}